# Shaping contactless forces through anomalous acoustic scattering

M. Stein[1], S. Keller[1], Y. Luo[1], O. Ilic[1]*

[1]Department of Mechanical Engineering, University of Minnesota, Minneapolis, 55455, USA

*Corresponding author. Email: ilic@umn.edu

**Abstract:** Waves impart momentum and exert force on obstacles in their path. The transfer of wave momentum is a fundamental mechanism for contactless manipulation, yet the rules of conventional scattering intrinsically limit the radiation force based on the shape and the size of the manipulated object. Here, we show that this intrinsic limit can be overcome for acoustic waves with subwavelength-structured metasurfaces, where the force becomes controllable by the arrangement of surface features, independent of the object's overall shape and size. Harnessing such anomalous metasurface scattering, we demonstrate complex actuation phenomena: self-guidance, where a metasurface object is autonomously guided by an acoustic wave, and contactless pulling, where a metasurface object is pulled by the wave. Our results show that bringing metasurface physics, and its full arsenal of tools, to the domain of mechanical manipulation opens the door to diverse actuation mechanisms that are beyond the limits of traditional wave-matter interactions.

**INTRODUCTION**

The process of momentum exchange between a wave and an object fundamentally underpins radiation forces exerted over liquid, gaseous and solid matter [1]. For macroscale objects, applications of wave forces have led to actuation mechanisms such as acoustic handling and transport [2–8], spawning interest as promising tools across biology and biomedicine to chemistry and colloidal science [9–17]. However, to date, such mechanisms have relied on conventional laws of wave refraction that govern how a wave interacts with the object it manipulates. For a typical surface or interface, the transfer of momentum is governed by Snell's law for waves, which means that the wave force is dictated—and constrained by—the overall shape and size of the object. In turn, this reliance constrains the kind of forces that can be induced, and therefore limits the range of actuation behaviors that can be generated. For example, common actuation configurations work only for subwavelength objects [18–20] or objects of particular symmetric shape [21–23] where actuating energy potentials are easy to create. Further, efforts to overcome such limitations have relied on complex active feedback control, typically employing some form of adaptive adjustment of acoustic fields, often in conjunction with object tracking [24,25]. At its root, the challenge lies in the entangled relationship between the shape of the object (which dictates scattering) and its mechanical response to waves (which is dictated by scattering).

Here, we demonstrate that patterning the object's surface into a metasurface—an embedded array of subwavelength elements—can overcome such wave force limitations by means of localized phase control of wave refraction. Metasurfaces are attracting significant interest due to their remarkable capability to synthesize complex wavefronts [26–33], with diverse applications in beam-steering, acoustic cloaking, and sound focusing [34–43]. Similar to optical manipulation [44–46], the ability to steer acoustic waves can also be seen as a mechanical phenomenon, and the underlying metasurface physics can be harnessed to shape mechanical forces. Unlike conventional wave refraction, a metasurface object exhibits anomalous refraction that is expressed through the generalized Snell's law, $\sin\theta_R = \sin\theta_I + \frac{1}{k}\frac{d\Phi(\vec{r})}{d\vec{r}}$, where the relationship between the refracted angle ($\theta_R$) and the incident angle ($\theta_I$) is augmented by the local phase term $\Phi(\vec{r})$ at the position $\vec{r}$, as shown in **Figure 1** (here, $k$ is the wave-vector). Since the relationship between the refracted and incident angle reflects the balance between the outgoing and the incoming wave momentum, the control of the local phase $\Phi(\vec{r})$ thus translates into the control of the local force $\vec{F}(\vec{r})$ that is exerted on the metasurface object. Therefore, the purposeful arrangement of metasurface unit cells on the subwavelength scale waves becomes a new way to realize desired actuation behavior on the large (object) scale.

We focus on metasurfaces as promising platforms for shaping mechanical actuation because they can endow an object with acousto-mechanical functionality that is independent of, and unconstrainted by, its morphology or dimensions. To demonstrate this capability, we synthesize three metasurface objects that are seemingly identical (all are rectangular blocks) but exhibit three different actuation functionalities. First, we show a metasurface whose subwavelength unit cells are arranged to enable strong lateral wave momentum shift, leading to a dominant sideways force that is *parallel* to the object surface. This stands in contrast to conventional refraction, where the only force that is exerted is *normal* to the surface. Second, we show a metasurface with self-guiding capability: it can sense the change in the acoustic field around it and respond to the change in an autonomous manner (**Figure 1b**). Specifically, the metasurface is able to lock itself to the remote acoustic source: when the source is moved, the metasurface object follows. Third, we show tractor beam-like behavior where a metasurface is contactlessly pulled towards the source of radiation (**Figure 1c**). It is noteworthy to point out that all three of these phenomena (a strong lateral force, self-guiding, pulling) are typically not possible for such rectangular blocks, as conventional scattering cannot provide the necessary momentum imbalance. But here such shape constraints are removed – in fact, apart from the different pattern of subwavelength features, the overall shape of the metasurface is the same in all cases in this work.

## RESULTS

To probe the connection between the subwavelength arrangement of metasurface unit cells and the radiation pressure it experiences, we first establish a predictive metasurface force model. Intuitively, a metasurface that steers sound waves in a particular direction should experience a force opposite to the change in the wave momentum $\Delta\vec{k}$ (**Figure 1**). To obtain a structure-to-force mapping that will guide our metasurface design, we employ a computational approach to calculate the second-order pressure terms. We numerically estimate the radiation force on a metasurface as the closed contour integral of the second-order acoustic components, i.e., $\vec{F} = -\int_S [\langle p_2 \rangle \hat{n} + \rho_0 \langle (\hat{n} \cdot \vec{v}_1)\vec{v}_1 \rangle]dA$, where $v_1, p_2$ are the first and second order perturbations in velocity and pressure, respectively, and $\rho_0$ is the density of the medium [47,48]. The integral can be carried out along any fixed surface $S$ enclosing an object, as ensured by momentum conservation. We validate this approach (Supplementary Materials [49]) and use it to infer the direction and the magnitude of the acoustic force on a metasurface. Importantly, we use this framework to guide the design of the metasurface—that is to select the arrangement of its unit cells—for each of the three target mechanical applications discussed in this work.

To demonstrate a proof-of-concept of contactless actuation of a metasurface, our first target mechanical behavior is a design that exhibits a strong lateral force. We select an inaudible operating frequency (20 kHz) for noiseless demonstration and to minimize disruptions. **Figure 2a** shows the building-block unit cell consisting of a U-shaped structure of subwavelength width. The depth of the grooves ($x_i$) determines the local scattered phase, and we map the relationship between the depth and the phase through finite element simulations of acoustic propagation in COMSOL Multiphysics (Fig. S3). Arranging the unit cells based on the depth-phase relationship, we can achieve a sideways wavefront shift and, further, we can refine the topology of the metasurface to enhance the lateral force. Specifically, we can treat the depths of the $N$ unit cells $(x_1, ..., x_N)$ as varying, but independent design parameters for an iterative optimization process where at each step $F(x_1, ..., x_N)$ is evaluated through finite-element analysis.

The outcome of this process is a metasurface shown in **Figure 2a (bottom)**, comprising of N = 30 unit cells. For scale, a US dime is also shown. The description of the metasurface and its dimensions is presented in the Supplementary Materials [49]. **Figure 2b** shows the scattered wave profile obtained from finite-element analysis in Comsol Multiphysics. Radiation is normally incident on the metasurface from a set of piezoelectric transducers (operating at $f$=20 kHz, $\lambda$=17 mm), to mimic the experimental configuration which is discussed below. We observe strong asymmetric scattering to the right, which is a signature that the metasurface should be experiencing a lateral push to the left. Interestingly, the side-scattering, while strong, exhibits an angular spread. In a conventional steering/focusing application, such an angular spread might be detrimental and undesirable in a metasurface. But for mechanical actuation, the key metric is the force; as long as the desired force profile is maintained, it is useful that a metasurface can in fact tolerate and be robust to angular spreading in scattering.

Our proof-of-concept experimental demonstration of metasurface actuation is displayed in **Figure 2c,d**. In the experiment, the metasurface is oriented with its unit cells facing a set of piezoelectric transducers operating at 20 kHz. The setup developed to measure actuation due to acoustic waves is based on the torsion-pendulum concept (details in the SI, e.g., Fig. S1 for the rendition of the setup). When a wave anomalously refracts off of a metasurface, a lateral force is induced, and the pendulum deflects in response. This motion is observed as the displacement of the laser spot from a laser that is reflected off the mirror at the pendulum base (**Fig. 2c**). A camera feed of the screen provides live tracking of metasurface actuation. Simultaneously, a spot-tracing algorithm detects the laser spot in real time and translates this information into the position of the metasurface. Initially, the metasurface is not actuated. As soon as sources are turned on, a sharp jump in the equilibrium position is observed (**Fig. 2e**, dashed red line labeled "on"); once the sources of sound are turned off, the metasurface returns to its initial equilibrium position. See Movie S1 for details. The photo in Fig. 2d shows the side view of the actuated metasurface. Crucially, the direction of the movement aligns with the predicted direction of the radiation force, indicating that the metasurface is indeed actuated as designed.

We characterize the strength of the actuation as a function of the signal supplied to the transducers. **Figure 2f** displays the change in the metasurface equilibrium position when the voltage is varied. A theoretical model predicts that the displacement will scale with the force as described by the relationship $F = \frac{\eta R^4 \pi}{2hL}\psi$, where $\psi$ is the torsional rotation angle, $R$ is the radius of the tungsten fiber, $\eta$ is the shear modulus of the fiber, and $h, L$ are the length of the fiber and the length of the pendulum arm, respectively [49]. The data in Figure 2f confirm the trend of stronger force with increased transducer voltage, but also show a deviation from the expected behavior that is more pronounced at higher source voltages/powers. We primarily attribute this discrepancy to the fact that the metasurface, when actuated, is noticeably displaced relative to the acoustic source and thus not intercepting the full acoustic wave at normal incidence – as can be seen in Fig. 2d. An additional source of discrepancy is the intrinsic variability of phase and intensity profile among the transducers. In the experiment, the orientation of the source, the metasurface, and the rotating arm are intentionally chosen to ensure that no parasitic forces can affect deflection (see Fig. S1). Finally, a measurement with a flipped metasurface yields the same behavior but with the opposite sign of deflection, eliminating the possibility that the observed effect is an experimental artifact, and further confirming that the origin of the force is the metasurface itself.

Having established the proof-of-concept of metasurface actuation, we proceed to introduce and demonstrate the self-guiding mechanism. A metasurface that guides itself must be capable of responding to changes in the acoustic intensity in its vicinity. **Figure 3a** shows the configuration where the source—comprising a set of transducers—is allowed to freely move. The curved orientation of the transducers is chosen to define a radiation axis that sets the equilibrium position of the metasurface. To probe the self-guiding phenomenon, we design a metasurface with center-symmetric arrangement of unit cells. When the metasurface is aligned with the axis of the radiation source, the scattering is symmetric, and no lateral force is present. However, when the source is offset, a net force is induced due to the asymmetry in scattering. The fabricated metasurface, shown in Figure 3a (right), is designed to drive itself back to its on-axis equilibrium position. For scale, a US dime is also shown. The description of the metasurface and the transducer arrangement is presented in the Supplementary Materials [49].

**Figure 3b** displays the simulated acoustic intensity profile without (top) and with (bottom) the metasurface in the computational domain. In the absence of the metasurface, the radiation source exhibits a focused profile with a defined radiation axis (solid line, Fig. 3b top). When the metasurface is introduced and the source is offset to the right, we observe strong refraction sideways to the left. This indicates that the metasurface would experience a force in the opposite direction (to the right), thus tracking the source. **Figure 3c** plots the simulated lateral force on the metasurface as a function of the source position. The force is normalized to the maximum force achieved at the source displacement of $\approx 25$ mm. The shape of the force curve confirms guiding behavior: the force on the metasurface is positive for positive source position, and vice-versa.

To experimentally demonstrate the self-guiding capability, we characterize the motion of the metasurface in response to a moving source. **Figure 3d** displays the pre-programmed position of the radiation source as a function of time (red dashed line), and the corresponding measured position of the metasurface (blue solid line). We indeed observe that the metasurface is autonomously guided towards the moving source, as indicated by the correlated trends of the two lines (red and blue). This self-guiding behavior is further highlighted by Movie S2. **Figure 3e** shows the top-view photos for three selected time instances: when the radiation source is aligned with the metasurface (middle panel), when the source is horizontally offset to the left by 15° (left panel), and when the radiation source is offset to the right by 15° (right panel). We comment that the actual dynamics of self-guiding will depend both on the metasurface design as well as the profile of the field intensity gradient away from the radiation axis. These two design features provide direct control over the strength of the guiding effect and can be further tuned depending on the target use. Notably, these results demonstrate metasurface self-guiding along one axis, but the concept is applicable to guiding along multiple directions with an appropriately designed metasurface.

Finally, we demonstrate the mechanism of attraction, where a metasurface is contactlessly pulled in the direction towards the radiation source. As indicated in the schematic in Fig. 1, an object will be pulled if it can efficiently forward scatter the incident wavefront to provide it with additional lateral momentum. Here, we synthesize a metasurface that feels an attractive/pulling force from a source positioned on one side, as shown in **Figure 4a.** The fabricated metasurface is shown in Fig. 4a (right) and described in the Supplementary Materials [49]. For scale, a US dime is also shown. **Figure 4b** plots the simulated lateral force on the metasurface as a function of the source angle, and reveals a region of negative values which denote an attractive force. In the experiment, we measure the displacement of the metasurface as a function of the attack angle of the source $\theta$. In our experimental configuration, a negative metasurface displacement corresponds to the presence of an attractive force. This is indicated by a shaded region in **Figure 4d**. For the metasurface in our case, we observe the force is attractive for a range of source attack angles, with the transition from pulling to pushing (negative to positive deflection) occurring at approximately 22.5 degrees. For attack angles greater than the transition cutoff, the force/deflection becomes positive. The observed metasurface deflection and the presence of the attractive force is consistent with the numerical predictions in Figure 4b. It should be noted that the simulation does not mimic the exact experimental situation – the simulation is a two-dimensional assessment of the force on a still/static metasurface, whereas in the experiment the metasurface freely moves in response to the experienced force. We remark that the acoustic 'pulling' result presented here is primarily indented to showcase the potential of metasurfaces to shape contactless forces; it is likely that stronger pulling behavior could be achieved in more advanced metasurfaces or in metasurfaces whose design optimization is guided by full three-dimensional wave simulations (as opposed to the limited 2D analysis employed in this work for convenience).

Conceptually, this demonstration shows that an attractive force can be realized on an object whose shape would otherwise not allow this possibility (here, the overall object shape is that of a rectangular slab). This is in contrast to conventional scattering, where attractive force requires the momentum imbalance that can only be achieved in objects of specific conical/tapered shape (e.g., a tapered prism [22]) which can constrain the overall functionality of the object. Metasurfaces, on the other hand, are not subject to such stringent shape limitations. The metasurface presented here operates in the reflection mode, but a transmissive metasurface could offer a similarly compelling tractor beam capability (as schematically shown in **Fig. 1d** for two independent and non-interacting incident beams).

**SUMMARY & OUTLOOK**

Our results demonstrate how the modes of contactless acoustic actuation can be tailored by the deliberate subwavelength patterning of the object's surface. Among our main findings, we show that the arrangement of subwavelength surface features can become a powerful design degree of freedom to control the radiation force, independent of the object's shape and size. We further harness this idea to demonstrate example mechanisms of metasurface self-guiding and metasurface tractor beaming. All actuation mechanisms in this work utilized the same block-shaped metasurfaces with the same unit cell shape, which further points to the versatility of the proposed approach. For future studies, it will be interesting to probe actuation dynamics that can be afforded by more complex metasurface morphologies and unit-cell topologies, in both the reflection and transmission modes of operation.

We envision a number of compelling extensions of this work. Here, we focused on structures with surface variability along only one direction, and it would be relevant to extend this idea to metasurfaces designed for actuation along multiple axes (e.g., multi-axial guiding). In particular, this could pave the way for concepts such as metasurface levitation, with the potential to overcome the constraints of conventional levitation. In typical levitation schemes, an object is stabilized and trapped by the acoustic intensity gradient, which works for objects small enough to fit into the nodes of the pressure wave (i.e., smaller than the wavelength). In contrast, it is possible to envision a metasurface able to generate its "own" trapping potential to stabilize itself in space, without relying on the profile of the acoustic field. In our demonstrations, the selection of the acoustic frequency beyond the human hearing range has potential benefits for noiseless operation and minimizing disruptions. However, we note that the presented concept can be straightforwardly scaled for actuation at other frequencies. Finally, the shaping of forces in soft and flexible materials, as well as non-reciprocal materials, are relevant directions for future research [50–52].

On a conceptual level, the ability to spatially tailor the radiation pressure with subwavelength precision enables the acousto-mechanical response of an object to be designed and realized independently of its dimensions and shape. This feature could be particularly compelling for endowing the object with additional, non-acoustic functionality (e.g., electronic, optical, magnetic). These demonstrations, combined with the ability to create sophisticated acoustic backgrounds and complex metasurface profiles, highlight the potentially diverse space of mechanisms for contactless actuation shaped by metasurface physics.

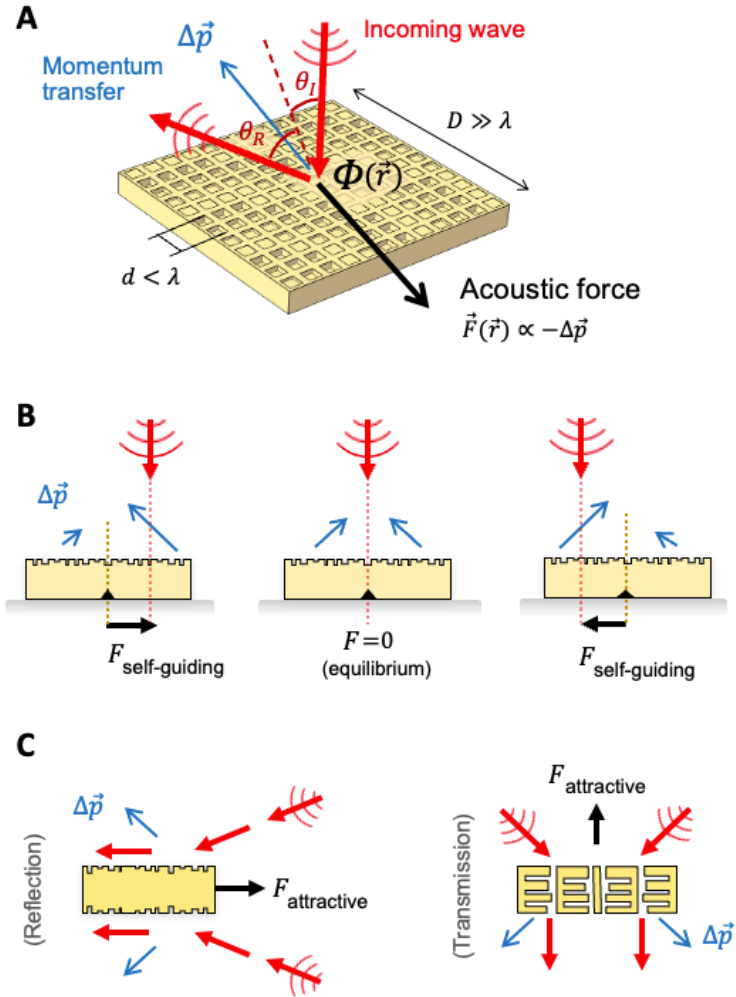

**Fig 1. Shaping contactless forces with metasurfaces that steer acoustic waves.** (**A**) The local force $\vec{F}(\vec{r})$ is proportional to the momentum change $\Delta\vec{p}$ (blue), which is controlled by the local phase $\Phi(\vec{r})$. Through spatial design of the subwavelength ($d < \lambda$) unit cells of the metasurface, complex force profiles can be generated to enable novel mechanisms for contactless actuation: (**B**) Self-guiding metasurfaces can autonomously self-lock to the remote wave source. Their ability to move as the source moves stems from the scattering asymmetry which pins the center of the metasurface to the source radiation axis (dashed red). E.g., when the metasurface is aligned on axis, no net force is present (middle); but if the source is moved the metasurface will follow (left, right). (**C**) Pulling force: metasurfaces can be not just pushed from, but also *pulled* towards the radiation source (here shown as two independent/noninterfering incident waves).

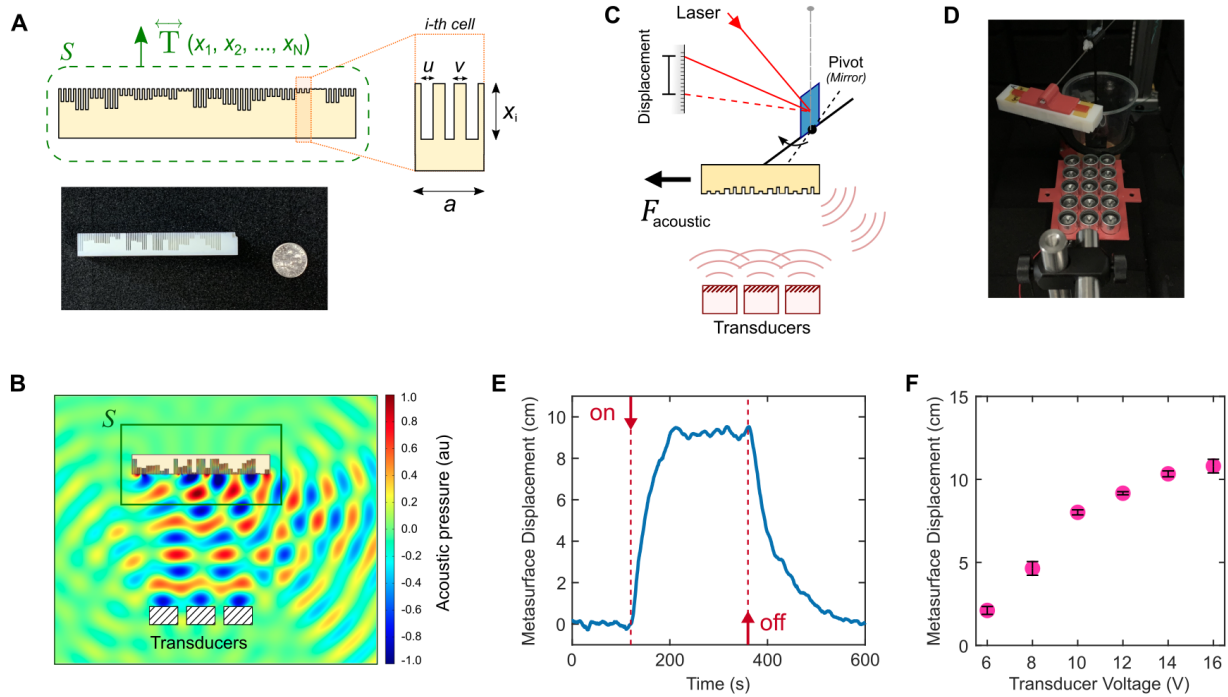

**Fig. 2. Design of the acousto-mechanical metasurface and the demonstration of actuation.**
(**A**) The building-block unit cell with variable depth (inset). Metasurface is optimized to induce a strong *lateral* force parallel to its long side, as derived from the pressure tensor $T$. Below: fabricated metasurface (a dime is shown for scale). (**B**) Scattering profile for the incident inaudible acoustic wave (20 kHz) showing the momentum flow to the right. Box indicates the enclosing surface over which the pressure tensor is numerically integrated. (**C**) Schematic of the setup to measure metasurface actuation (see Fig. S1 for details). (**D**) Photo of the metasurface when actuated (Movie S1). (**E**) Metasurface displacement versus time. Dashed lines denote the turning on/off of the sound source. (**F**) Metasurface equilibrium displacement versus transducer voltage.

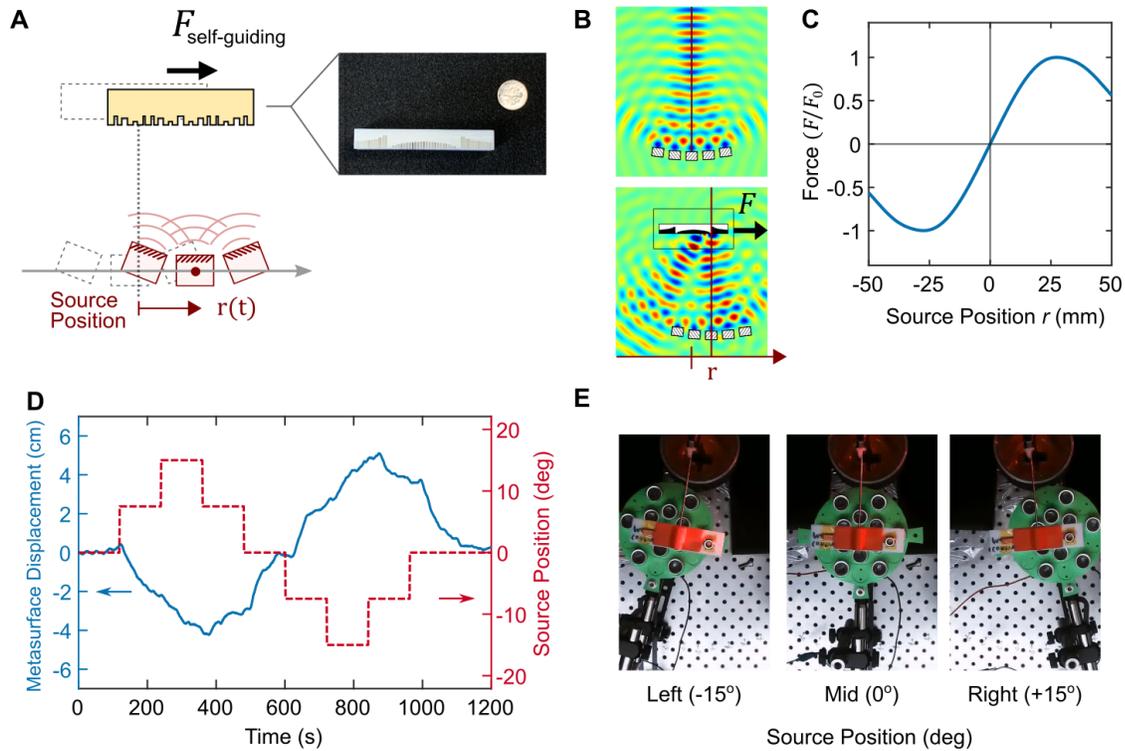

**Fig. 3. Metasurface self-guiding: concept and demonstration.** (**A**) Metasurface is designed to autonomously track and guide itself to follow the movement of the radiation source. Right: fabricated metasurface (a dime is shown for scale). (**B**) Top: Focused acoustic intensity profile emanating from a set of piezoelectric transducers that comprise the source of radiation (20 kHz). The solid line denotes the radiation axis. Bottom: Scattering map for a case when the metasurface is displaced away from the radiation axis. (**C**) Force as a function of the source position. The shape of the curve confirms guiding behavior: force is positive for positive source position, and vice-versa. (B,C) are simulation results. (**D**) Experiment: the metasurface position (blue) tracks the position of the radiation source (red), as a function of time. (**D**) Top view of the camera frames at several instances during actuation (Movie S2).

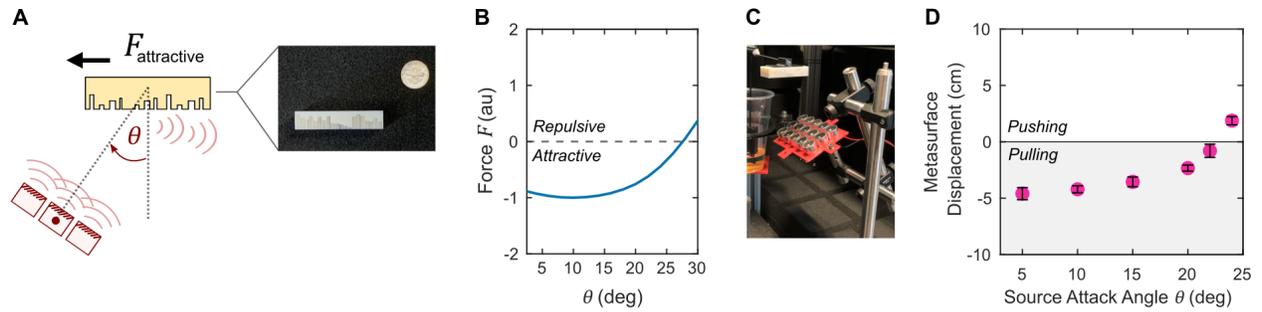

**Fig. 4. Metasurface attractive forces: concept and demonstration.** (**A**) Shown is the configuration of a metasurface designed to experience a force that attracts it towards the radiation source. Right: fabricated metasurface (a dime is shown for scale). (**B**) Lateral force as a function of the source attack angle. Negative values denote attractive force. The force is extracted from simulation and normalized to its peak negative magnitude (~10 deg). (**C**) Photo of the metasurface in the setup. (**D**) Experimentally measured metasurface displacement as a function of the source attack angle $\theta$, as defined in (A). Shaded area indicates the condition where a metasurface is pulled towards the source.